\documentclass{pazh2col-utf}

\usepackage{graphicx}
\usepackage{amsmath}
\usepackage{amssymb}
\usepackage{color}

\newcommand{\vect}[1]{\boldsymbol{#1}}
\newcommand{\kms}{\,\,km\,s$^{-1}$\,}
\newcommand{\cmc}{\,\,cm$^{-3}$\,}

\newcommand{\ergs}{\,\,erg\,s$^{-1}$\,}
\newcommand{\Msun}{\,\,M$_{\odot}$\,}

\newcommand{\MsunYr}{\,\,M$_{\odot}$\,yr$^{-1}$\,}

\usepackage[compact]{titlesec}
\titlespacing{\section}{0pt}{3ex}{2ex}
\titlespacing{\subsection}{0pt}{2ex}{2ex}
\begin{document}

\renewcommand{\figurename}{{\small\textbf{Fig.}}}
\renewcommand{\tablename}{{\small\textbf{Table}}}

\def\spislit{\small REFERENCES}
\def\emailname{e-mail}
\def\keywordsname{Keywords}

\authors[BADMAEV et al.]{
  \nextauth[danirbadmaev.astro@gmail.com]{\hspace{-4mm}D.~V. Badmaev}{1},
  \nextauth{A.~M. Bykov}{1},
  \nextauth{M.~E. Kalyashova}{1}
}

\titles[Shells and magnetized envelopes]{Shells and bubbles around compact clusters of massive stars: 3D MHD simulations}

\affiliations{
\nextaffil{Ioffe Institute, 26 Politekhnicheskaya St, 194021, Saint Petersburg, Russia}
}

\def\postupila{\vspace{5mm}Received 16.10.2025 \\
Revised 23.10.2025; accepted 23.10.2025}

\def\journame{ASTRONOMY LETTERS ~ Vol. 51 ~ No. 6 ~ 2025}

\wideabstract{We present the results of three-dimensional magnetohydrodynamic (3D MHD) simulations of the plasma flow structure in the vicinity of a compact cluster of young massive stars. The cluster is considered at the evolutionary stage dominated by Wolf-Rayet stars. This stage occurs in clusters with ages of several million years, close to the onset of supernova explosions; the well-known objects Westerlund 1 and 2 are the prototypes. The collisions of powerful winds from massive stars in the cluster core, calculated as interactions of individual outflows, are accompanied by their partial thermalization and produce a collective cluster wind. The MHD dynamics of the cluster wind bubble expansion into the interstellar medium is considered, depending on the density of the surrounding medium with a uniform magnetic field. We show that when expanding into a cold neutral medium, the cluster wind is able to reshape its surrounding environment over the Wolf-Rayet star lifetime, sweeping up more than $10^4$\Msun of gas in $\sim2\times10^5$ yr and producing extended, thin and dense shells with an amplified magnetic field. In a cold neutral medium with a density of $\sim20$\cmc and a magnetic field of $\sim 3.5$ $\mu$G, a thin shell forms around the cluster wind bubble, characterized by a cellular structure in its density and magnetic field distributions. The cellular magnetic field structure appears in parts of the shell expanding transversely to the orientation of the external magnetic field. Magnetic fields in the shell are amplified to strengths $\gtrsim 50$ $\mu$G. The formation of the cellular structure is associated with the development of instabilities. The expansion of the bubble into a warm neutral interstellar medium also leads to the formation of a shell with an amplified magnetic field.
\doi{10.1134/S...}
}

%=========================================================

\section{Introduction}
\label{introduction}
Powerful stellar winds interacting with the interstellar medium (ISM) can form extended bubbles filled with wind material and bounded by a dense gaseous shell (Pikelner 1968; Avedisova 1972; Castor, McCray, Weaver 1975; Weaver et al. 1977). At the same time, a significant fraction of massive stars are born in compact young massive star clusters (YMSCs) or extended OB associations (Lada \& Lada 2003; de Wit et al. 2005; Adamo et al. 2020). Powerful stellar winds in massive star clusters shape their local environment. At later evolutionary stages of such systems, the collective action of stellar winds and supernova explosions from collapsing massive stars sweeps out material around the cluster, creating so-called superbubbles ranging in size from tens of parsecs to kiloparsecs, filled with hot, rarefied, turbulent gas and surrounded by massive gaseous shells (McCray \& Kafatos 1987; Mac Low \& McCray 1988; Cox 2005). Examples of such objects include the Cygnus superbubble, the Orion-Eridanus superbubble surrounding the Orion OB1 association, the Vela superbubble around the Vela OB2 association, and the giant 30 Doradus bubble in the Large Magellanic Cloud. Star-forming regions in galaxies, containing multiple clusters of young massive stars, drive powerful outflows into galactic halos (Chevalier \& Clegg 1985; Canto et al. 2000). Galactic outflows from star-forming regions may be associated with large-scale structures observed in radio and X-ray surveys, such as giant bubbles (eROSITA bubbles) (Predehl et al. 2020) and the North Polar Spur (Churazov et al. 2024).

Compact clusters of young massive stars are abundant in galaxies with active star formation. Observations with the JWST NIRCam F250M filter of the central kiloparsec of the nearby galaxy M82, hosting a powerful galactic wind outflow, allowed Levi et al. (2024) to identify 1357 candidate young star clusters with masses above $10^4$\Msun and a median cluster radius of about 1 pc. Analysis of CO and H$_{\alpha}$ emission observations of giant molecular clouds in nine nearby galaxies, performed by Chevance et al. (2022) with an angular resolution of about one arcsecond, estimated the cloud destruction time as $\sim3$ Myr after the appearance of massive stars. This short timescale indicates the important role of massive star winds and somewhat changes the notion of the dominant role of supernovae in this process (see the review by Schinnerer \& Leroy 2024). Thus, modeling the dynamics of compact massive star clusters at the stage preceding supernova explosions is an essential element in constructing a consistent theory of active star-forming regions. In addition to powerful ionizing UV radiation from stars, a factor influencing the surrounding environment on scales of $\sim10$ pc are supersonic MHD flows of plasma heated to X-ray temperatures from the region of colliding fast winds in the core of a compact cluster, as well as a broad spectrum of non-thermal particles capable of penetrating dark dense clouds and maintaining certain heating and ionization rates necessary for star formation processes.

The detection of high-energy gamma-ray emission in the direction of several YMSCs (Abramowski et al. 2012; Yang et al. 2018; Ackermann et al. 2011; Cao et al. 2021) clearly indicates that, alongside supernovae, they are accelerators of high-energy cosmic rays. In some cases (e.g., for the Cygnus Cocoon and the gamma-ray source around the compact cluster Westerlund 1), the gamma-ray emission occupies a spatial region of tens of parsecs, corresponding to a superbubble (Aharonian et al. 2019). This highlights the need to study the immediate surroundings of the actual particle accelerators – star clusters.

Magnetic fields play a key role in cosmic ray acceleration and generation of non-thermal radiation, directly influencing particle confinement. In colliding flow systems, such as massive star clusters, significant amplification of the turbulent magnetic field can occur, where part of the kinetic energy of stellar winds and/or supernova remnant shocks converts into magnetic energy. An intermittent, highly turbulent medium with strong primary and weak secondary shocks forms, covering a broad spectrum of MHD fluctuations. Previously, as part of investigating turbulence in YMSCs, 3D MHD modeling of multiple interacting stellar winds in the cluster core was performed (Badmaev et al. 2022). It was shown that magnetic fields in such a system are highly intermittent and amplified to $\sim300$ $\mu$G due to the energetic winds of young massive stars. In a subsequent study (Badmaev et al. 2024), a supernova explosion in the core of a YMSC was simulated, and perturbations in density, velocity, temperature and magnetic field resulting from the explosion were studied. The relaxation time of the cluster to its pre-supernova state was found to be $\lesssim10$ kyr. Thus, using MHD modeling, all main characteristics of a YMSC core, as well as their temporal evolution, and the impact of the supernova on the cluster medium were investigated.

The necessary next step is to study the interaction of the cluster with the external space – the interstellar medium. External conditions, such as the density and size of the parent molecular cloud and the configuration of the external magnetic field, should significantly influence the formation and dynamics of the collective cluster wind, which carries the total kinetic energy of the winds of individual cluster member stars. The collision of the stellar wind with a dense, magnetized external medium could potentially lead to local magnetic field amplification, creating conditions favorable for particle confinement and acceleration in this region.

Modeling of the plasma flow structure in star clusters using the smoothed particle hydrodynamics method was performed by Rockefeller et al. (2005) for the YMSCs Arches and Quintuplet. The influence of different spatial stellar distributions within a cluster on the formation of a collective wind was studied by Rodriguez-Gonzalez et al. (2007). The problem of forming dense filamentary structures within the cluster volume under optically thin cooling was considered by Rodriguez-Gonzalez et al. (2008). A hydrodynamic (HD) model of the interaction of a collective wind from a cluster of several massive stars with an inhomogeneous parent molecular cloud was proposed by Rogers \& Pittard (2013). Gupta et al. (2018, 2020) used a 3D HD model of a cluster wind propagating into a uniform ISM to study non-thermal emission from YMSCs and the role of the cluster wind termination shock in cosmic ray acceleration. The role of turbulence induced by colliding stellar winds within the cluster was considered by Gallegos-Garcia et al. (2020). Härer et al. (2024) presented a test model of a cluster wind bubble of a YMSC in a dense medium, performed in ideal MHD with a coarse stretched grid.

In the work of Vieu et al. (2024), HD modeling of a sparse star cluster reproducing the characteristics of the Cygnus OB2 association was performed. The modeling showed that stars in such a cluster are too far apart to form a cluster wind termination shock. Härer et al. (2025) performed 3D MHD modeling of the collective wind from a compact ($\lesssim 1$ pc) cluster of massive stars and investigated the morphology of the resulting flows, magnetic fields, and the termination shock. The modeling showed that, as expected for clusters containing dozens of stars with different kinetic powers, the cluster wind termination shock is not strictly spherical. According to the authors' estimates, the magnetic field values near the cluster wind termination shock do not allow confining and accelerating particles with energies $\sim1$~PeV.

In global galaxy models, in their central regions with active star formation up to 500 pc in size, magnetic field amplification to several tens of $\mu$G over timescales of about 250 million years is possible, as demonstrated by Moon et al. (2023).

%------------------------------------------------------
\section{Model}
\subsection{General Model Description}
It is assumed that in a YMSC with an age of $\sim3$ Myr, just before the first core-collapse supernova (CCSN) explosions, Wolf-Rayet (WR) stars are the dominant sources of mechanical energy released into the surrounding environment via powerful stellar winds.
Comprehensive and realistic modeling of the environment in the immediate vicinity of a YMSC core is extremely complex and involves accounting for many factors present during the first few million years of cluster formation: inhomogeneity of the parent molecular cloud, continuance of star formation burst, stellar initial mass function (IMF) distribution, feedback from the first massive stars and destruction of the parent cloud, evolution of massive stars, and stellar dynamics. The characteristic lifetime of a massive star in the WR phase, as well as the dynamical crossing time of a star in the cluster, is $\sim10^5$ yr.
The 3D MHD model constructed in this study describes, on spatial scales up to 20 pc, the dynamics of the interaction of a YMSC with the interstellar medium over a period $t_{\rm model}\sim10^5$ yr before the beginning of the supernova epoch. The star cluster during the modeling period has a given stellar population with an O:WR number ratio $\sim10$ and a mechanical luminosity $\gtrsim 10^{38}$\ergs. Current observations of the rich compact cluster Westerlund 1 show an even higher number of Wolf-Rayet stars with O:WR $\sim5$. The model describes the interaction of multiple massive star winds in the core of a compact cluster, accounting for radiative properties and thermal conductivity of the plasma.
The model does not account for effects of ISM inhomogeneity related to the hierarchical structure of molecular clouds with clumps of molecular gas ($n_{\rm ism} > 100$~cm$^{-3}$) near the cluster. Also, previous evolutionary stages of massive stars and associated mass loss preceding the modeling period are not considered.

\subsection{Numerical Scheme and Main Equations}
A detailed description of the numerical modeling of the YMSC core using the MHD code PLUTO (Mignone et al. 2007) is provided in Badmaev et al. (2022). The code integrates the following system of non-ideal single-fluid MHD equations:
\begin{gather}
    \frac{\partial\rho}{\partial{t}}+\vect{\nabla}\cdot\left(\rho\vect{u}\right)=0,\label{1}\\
    \frac{\partial\vect{m}}{\partial{t}}+\vect{\nabla}\cdot\left(\vect{m}\otimes\vect{u}-\vect{B}\otimes\vect{B}+\vect{\hat{I}}p_{\mathrm{t}}\right)=0,\label{2}\\
    \frac{\partial{E}}{\partial{t}}+\vect{\nabla}\cdot\left[\left(E+p_{\mathrm{t}}\right)\vect{u}-\vect{B}\left(\vect{u}\cdot\vect{B}\right)\right]=\Phi\left(T,\rho\right),\label{3}\\
    \frac{\partial\vect{B}}{\partial{t}}+\vect{\nabla}\cdot\left(\vect{u}\otimes\vect{B}-\vect{B}\otimes\vect{u}\right)=0,\label{4}
\end{gather}
where $\vect{m}=\rho\vect{u}$ is the momentum density, $\vect{B}$ is the magnetic field strength (normalized in the code by $\sqrt{4\pi}$), $\vect{\hat{I}}$ is the identity tensor, $p_{\mathrm{t}}=p+\vect{B}\cdot\vect{B}/2$ is the total pressure, $p$ and $\rho$ are the gas pressure and density, respectively. The total energy density is written as,
\begin{equation}
    E=\frac{p}{\gamma-1}+\frac{\vect{m}\cdot\vect{m}}{2\rho}+\frac{\vect{B}\cdot\vect{B}}{2},
\end{equation}
where $\gamma=5/3$ is the adiabatic index of an ideal monatomic gas. The system of equations (\ref{1}--\ref{4}) is closed by the ideal gas equation of state:
\begin{equation}
p=\frac{\rho{k}_{\rm{B}}T}{\mu{m}_{\rm H}},
\end{equation}
where $\mu$ is the mean particle mass in the gas, $m_{\rm{H}}$ is the mass of a hydrogen atom. It is assumed that in the neutral interstellar medium $\mu=1.22$, while in the ionized wind material $\mu=0.61$.

The MHD equation system is solved using the Godunov scheme on a uniform 3D grid with the HLLD (Miyoshi \& Kusano, 2005) approximate Riemann solver. The numerical scheme is second-order accurate: a linear TVD reconstruction of physical quantities between adjacent grid cells with a van Leer limiter is applied, and time discretization is performed using a 2nd-order Runge-Kutta method. The time step for integration is determined by the standard Courant-Friedrichs-Lewy (CFL) condition, here set to $C_{\rm cfl}=0.3$. The solenoidal condition for the magnetic field ($\vect{\nabla}\cdot\vect{B}=0$) is maintained throughout the computational domain by the Hyperbolic Divergence Cleaning algorithm (Dedner et al. 2002).

%%%%%%%%%%%%%%%%%%%%%%%%%%%%%%%%%%%%%%%%%%%%%%%%%%%%%%%%%
\begin{figure*}[ht!]
\centering
\includegraphics[width=\columnwidth]{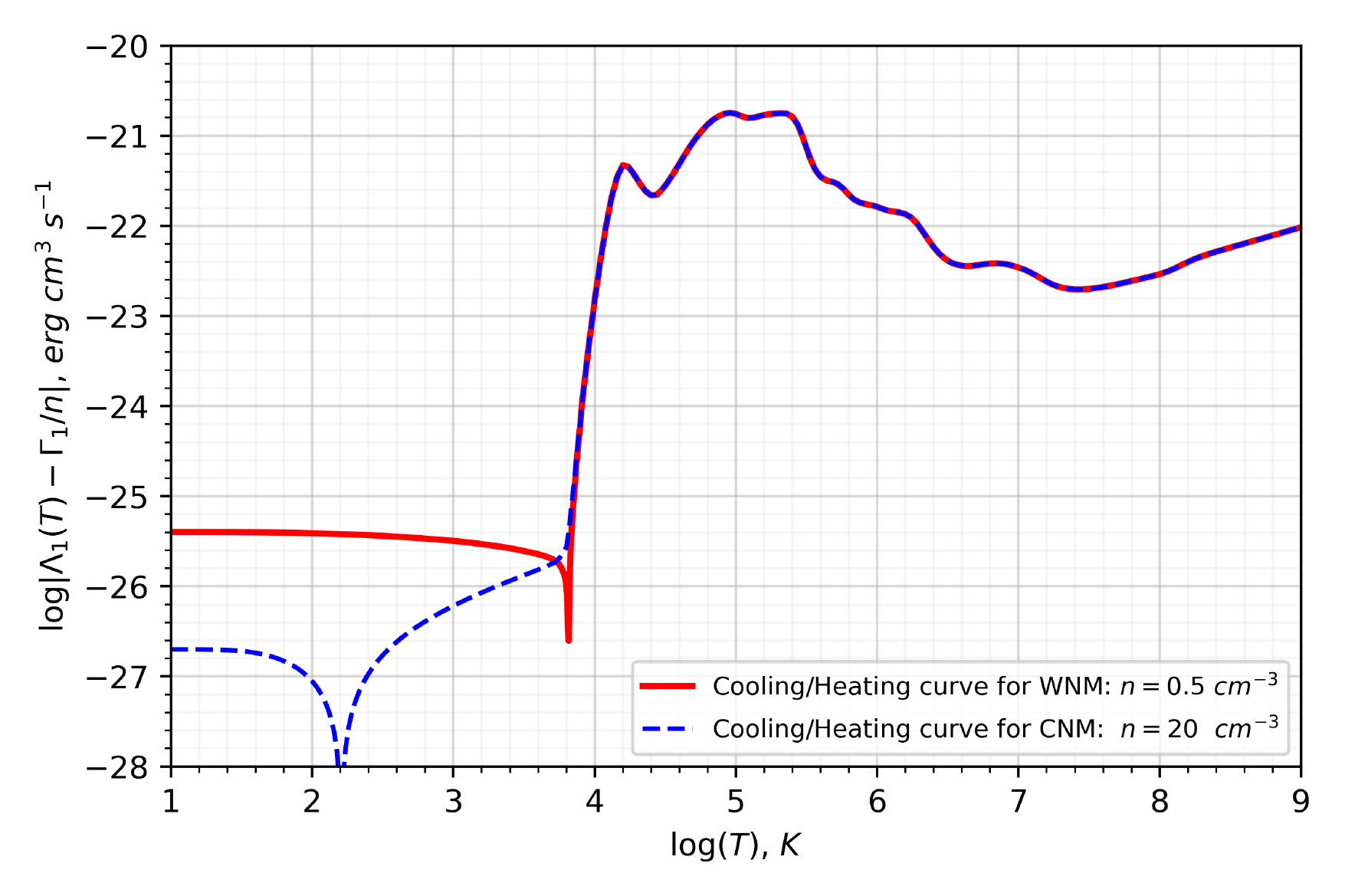}
\includegraphics[width=\columnwidth]{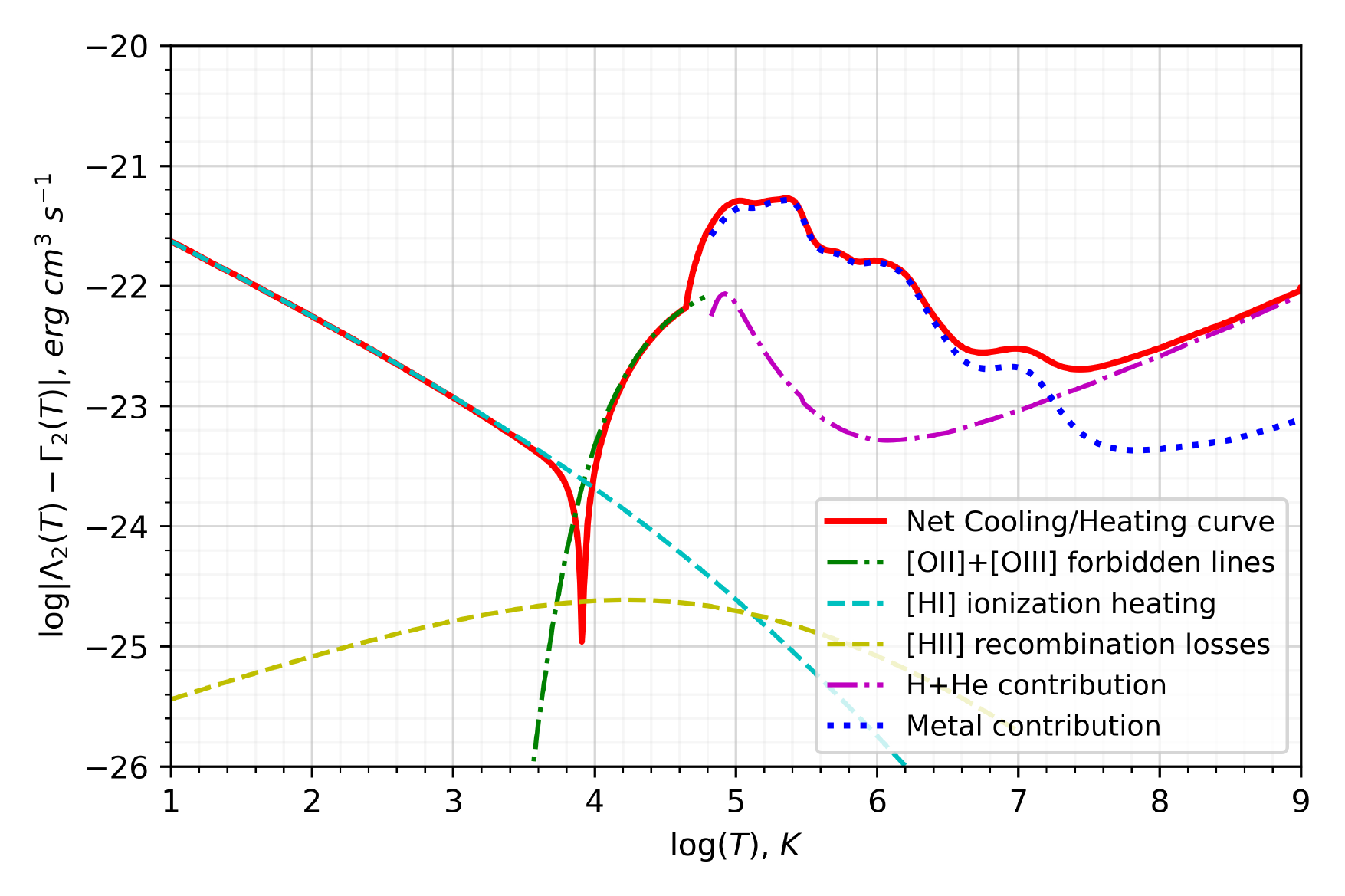}
\vspace{-1mm}
\caption{\textit{Left:} Absolute value of the cooling/heating function for an optically thin plasma in the two interstellar medium phases, CNM and WNM. The equilibrium temperatures $T_{\rm cnm}=160$~K (blue dashed) and $T_{\rm wnm}=6400$~K (red solid) correspond to the inflection points where the functions change sign. \textit{Right:} Absolute value of the cooling/heating function for an optically thin, ionized medium (red solid) bounded by the bubble shell and filled with cluster wind material. Dashed curves show individual components (Meyer et al. 2014): contributions from hydrogen and helium line cooling (magenta dash-dotted); metal line cooling (blue dotted; Wiersma et al. 2009); hydrogen recombination (yellow dashed; Hummer 1994); forbidden metal lines (green dash-dotted; Henney et al. 2009); and hydrogen photoionization heating (cyan dashed). The equilibrium temperature is $T_{\rm ion}=8300$~K.}
\label{fig1}
\end{figure*}
%%%%%%%%%%%%%%%%%%%%%%%%%%%%%%%%%%%%%%%%%%%%%%%%%%%%%%%%%

\subsection{Optically Thin Cooling}\label{sec:cool}
The gains and losses of thermal energy due to radiative processes in optically thin plasma read
\begin{equation}
    \Phi\left(T,\rho\right)=n^{\alpha}\Gamma_{\alpha}\left(T\right)-n^{2}\Lambda_{\alpha}\left(T\right),
\end{equation}
where $\Lambda_{\alpha}\left(T\right)$ and $\Gamma_{\alpha}\left(T\right)$ characterize radiative cooling and heating of the optically thin plasma, respectively, and $n = \rho/(\mu m_{\rm H})$ is the particle number density. The case $\alpha=1$ corresponds to the collisional ionization equilibrium (CIE) regime, while $\alpha=2$ corresponds to the photoionization equilibrium (PIE) regime in the plasma. The cooling function $\Lambda_{1}\left(T\right)$ in the temperature range $T<10^{4}$~K is approximated by formula (4)\footnote{The formula contains typographical errors corrected in Vázquez-Semadeni et al. (2007).} from Koyama \& Inutsuka (2002). The middle part of the cooling function, in the range $10^4<T<10^8$~K, is interpolated from Table~2 in Schure et al. (2009). For the high-temperature bremsstrahlung part, for $T>10^{8}$~K, the piecewise formula (22) from Schneider \& Robertson (2018) is used. The equilibrium temperatures of the interstellar medium, characteristic for CNM and WNM models, are achieved by including constant photoelectric heating (Wolfire et al. 2003; Klessen \& Glover 2016):
\begin{equation}
    \Gamma_{1}\left(T\right)=\Gamma_{1}=
    \begin{cases}
        2 \times 10^{-27}~~\text{erg s}^{-1}:&\text{CNM},\\
        4 \times 10^{-26}~~\text{erg s}^{-1}:&\text{WNM}.
    \end{cases}
\end{equation}
The resulting optically thin cooling/heating functions, $\Lambda_{1}(T)-\Gamma_{1}/n$, are shown on the left panel of Fig.~\ref{fig1} for the two cases of neutral ISM.

Within the bubble, the medium is assumed to be photoionized. The resulting cooling/heating function for ionized material, $\Lambda_{2}(T)-\Gamma_{2}(T)$, is constructed according to the recipe from Meyer et al. (2014), see the right panel of Fig.~\ref{fig1}. It is assumed that radiative heating $\Gamma_{2}(T)$ arises mainly from immediate photoionization of recombining hydrogen and is therefore a simple product of the recombination rate and the average heating energy per ionization:
\begin{equation}
    \Gamma_{2}(T)=\alpha_{\rm B}(T)\langle{E}_{\rm phi}\rangle~~\text{erg cm}^{3}~\text{s}^{-1},
\end{equation}
where $\alpha_{\rm B}(T)$ is the hydrogen recombination coefficient from Hummer (1994), $\langle{E}_{\rm phi}\rangle=5$~eV is the average kinetic energy of an electron emitted during hydrogen photoionization in the radiation field of a typical O-star (Osterbrock \& Ferland 2006; Green et al. 2019).

\subsection{Thermal Conduction}\label{sec:tc}
A separate calculation of the cluster wind bubble formation process was performed including thermal conduction. In this case, the energy balance equation~(\ref{3}) takes the form
\begin{align}
    \begin{split}
        \frac{\partial{E}}{\partial{t}}+\vect{\nabla}\cdot\left[\left(E+p_{\mathrm{t}}\right)\vect{u}-\vect{B}\left(\vect{u}\cdot\vect{B}\right)\right]=\\
        =\vect{\nabla}\cdot\vect{F}+\Phi\left(T,\rho\right)&,
    \end{split}
\end{align}
where, following Balbus \& McKee (1982),
\begin{equation}\label{Fcond}
    \vect{F}=\vect{F}_{\rm cl}\frac{F_{\rm sat}}{F_{\rm sat}+|\vect{F}_{\rm cl}|}.
\end{equation}
The standard anisotropic heat flux in magnetized plasma (Braginskii 1965):
\begin{equation}
    \vect{F}_{\rm cl}=\kappa_{||}\hat{\vect{b}}(\hat{\vect{b}}\cdot\vect{\nabla}T)+\kappa_{\perp}[\vect{\nabla}T-\hat{\vect{b}}(\hat{\vect{b}}\cdot\vect{\nabla}T)],
\end{equation}
where $\hat{\vect{b}}=\vect{B}/|\vect{B}|$ is the unit vector along the magnetic field lines. The thermal conductivity transverse to the magnetic field lines $\kappa_{\perp}$ is negligible compared to the longitudinal coefficient $\kappa_{||}$, which in the case of a neutral medium with low temperature, $T<6500$~K, follows Parker's (1953) estimate
\begin{equation}
    \kappa_{||}=2.5\times10^{3}T^{1/2}~~\text{erg c}^{-1}~\text{cm}^{-1}~\text{K}^{-1},
\end{equation}
while at high temperatures, in an ionized medium, it corresponds to Spitzer's (1962) estimate
\begin{equation}
    \kappa_{||}=5.6\times10^{-7}T^{5/2}~~\text{erg c}^{-1}~\text{cm}^{-1}~\text{K}^{-1}.
\end{equation}
The saturated heat flux (see Cowie \& McKee 1977) is determined by the formula
\begin{equation}
   F_{\rm sat}=-\frac{3}{2} \rho c^{3}_{\rm s,iso},
\end{equation}
where $c_{\rm s,iso}$ is the isothermal sound speed. Equation (\ref{Fcond}) allows a smooth transition between the classical and saturated thermal conduction regimes.

%%%%%%%%%%%%%%%%%%%%%%%%%%%%%%%%%%%%%%%%%%%%%%%%%%%%%%%%%
\begin{figure}[ht!]
\centering
\includegraphics[width=\columnwidth]{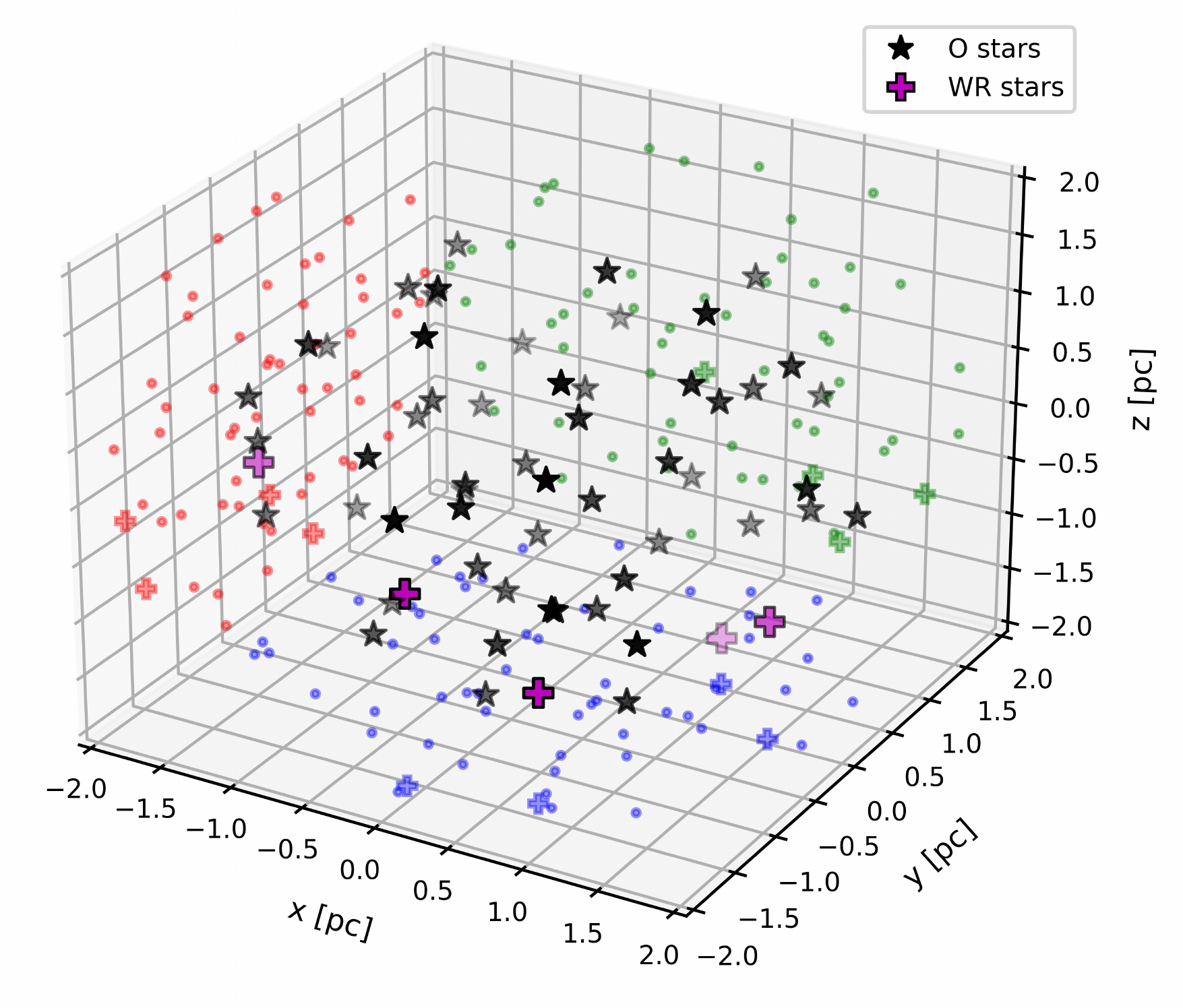}
\vspace{-5mm}
\caption{Positions of randomly distributed massive stars of different spectral classes within the computational domain representing the cluster core. Black asterisks mark O-stars; purple crosses mark WR stars. Marker opacity indicates line-of-sight depth. Red, green, and blue dots on the domain faces show projections of the stellar positions onto the \textit{yz, xz}, and \textit{xy} planes, respectively.}
\label{fig2}
\end{figure}
%%%%%%%%%%%%%%%%%%%%%%%%%%%%%%%%%%%%%%%%%%%%%%%%%%%%%%%%%

\subsection{Bubble Initialization}

\subsubsection{Cluster Wind}
The cluster consists of 60 young massive stars (55 O-stars and 5 WR stars), which generate a powerful collective wind. Stars in the cluster are randomly and uniformly distributed within a sphere of radius 2 pc, representing the YMSC core region, see Fig.~\ref{fig2}. Each star inside the computational domain represents a wind source which is implemented as a boundary condition (see Badmaev et al. 2022) within a small spherical region with a radius of 5 grid cells\footnote{And 3 grid cells while up-scaling, so that the region's physical size is $\lesssim0.2$~pc.}. O-star winds have the following parameters: velocity $v_{\rm o}=2400$\kms, mass loss rate $\dot{M}_{\rm o}=1.1\times10^{-6}$\MsunYr, mechanical luminosity $\dot{E}_{\rm o}=2\times10^{36}$\ergs. WR-star wind parameters: $v_{\rm wr}=1800$\kms, $\dot{M}_{\rm wr}=10^{-5}$\MsunYr, $\dot{E}_{\rm wr}=10^{37}$\ergs. Thus, the total mechanical luminosity of the cluster is $\dot{E}_{\rm tot}=1.6\times10^{38}$\ergs. The model does not include stars in the cold supergiant (CSG) phases, as their slow ($v_{\rm csg}\sim30$\kms) and dense ($\dot{M}_{\rm csg}\gtrsim10^{-4}$\MsunYr) winds do not contribute significantly to the mechanical luminosity, although they can be a substantial source of cool gas on a timescale $\gtrsim10^4$ yr (Badmaev et al. 2024). Detailed MHD modeling (Badmaev et al. 2022, 2024) indicates a significant, about 50\%, thermalization of kinetic energy in the YMSC core, resulting from the collision between fast O and WR winds. Magnetic fields carried by winds of massive stars ($\sim100$ G at stellar surfaces) are amplified to $\sim100$~$\mu$G in regions of strong compression, forming elongated filaments that leave the core region with flows of thermalized cluster wind. At the boundary of YMSC core, the cluster wind consists mainly of thermalized wind from stars located closer to the core center, and, locally, of powerful winds (especially WR) from stars located at the periphery.

\subsubsection{Interstellar Medium}
The model considers the interaction of the cluster wind with a uniform interstellar medium, represented in two characteristic phases: the cold neutral medium (CNM, $n=20$~cm$^{-3}$, $T=160$~K) and the warm neutral medium (WNM, $n=0.5$~cm$^{-3}$, $T=6400$~K). The pressure $nT=3200$~K~cm$^{-3}$ is the same for both phases. Interstellar gas uniformly fills the space within a 20 pc radius around the cluster center. The total gas mass in the computational domain volume is $\approx1.6\times10^{4}$\Msun for CNM and $\approx400$\Msun for WNM. The magnetic field within the distance of 20 pc around the cluster core can be considered quasi-uniform with a strength $B_{\rm ism}=3.5$~$\mu$G. More complex ISM configurations, including gas gradients and phase hierarchy, will be considered in future work.

The cluster wind and ISM material are tracked separately using a scalar marker $Q$, described by a linear advection equation:
\begin{equation}
    \frac{\partial\left(\rho{Q}\right)}{\partial{t}}+\vect{\nabla}\cdot\left(\vect{v}\rho{Q}\right)=0.
\end{equation}
The marker is passively advected with the physical flow and allows distinguishing between the cluster wind components and interstellar medium. For the ionized winds, values $Q>10$ are chosen, and for the neutral ISM gas $Q=1$. Separating material by origin allows applying two different plasma cooling/heating functions simultaneously.

\subsubsection{Scaling Method}
Due to the limited grid resolution and the high spatiotemporal dynamic range of the problem ($l_{\rm evol}\sim0.1-10$ pc, $t_{\rm evol}\sim10^5$~yr), a scaling method was employed. This method involves sequentially doubling the physical size of the cubic computational domain while keeping the number of cells in the 3D grid constant at $528^{3}$. The domain sizes were chosen so that at each scaling step, (1) $\rightarrow$ (2) $\rightarrow$ (3), spheres (bubbles) with radii of 5 pc $\rightarrow$ 10 pc $\rightarrow$ 20 pc could be inscribed within them with a small margin. Consequently, the full bubble calculation proceeds in three stages:
\begin{enumerate}
    \item The collective wind of the young massive star cluster (YMSC) and the associated bubble form via the merging of individual stellar winds. This stage is simulated in a domain with side lengths of $[-5, 5]$ pc and a spatial resolution of 0.02 pc cell$^{-1}$, filled with interstellar medium (CNM/WNM).
    \item The final state from the previous stage is mapped into a new domain with side lengths of $[-10, 10]$ pc. The simulation continues with the spatial resolution coarsened by a factor of two (to 0.04 pc cell$^{-1}$).
    \item The result is again mapped into a larger domain with side lengths of $[-20, 20]$ pc, reducing the initial resolution by a factor of four (to 0.08 pc cell$^{-1}$) for further evolution.
\end{enumerate}
This approach provides several key benefits: it allows the compact ($\sim1$ pc) YMSC core region to be resolved with sufficient accuracy at early times; it enables the model to be extended to longer timescales ($\gtrsim10^5$ yr) while retaining detail in intermediate results through gradual resolution reduction; it offers effective control over the computational cost (in core-hours); and it avoids geometric artifacts that can arise from the use of non-uniform or logarithmically stretched grids.

%%%%%%%%%%%%%%%%%%%%%%%%%%%%%%%%%%%%%%%%%%%%%%%%%%%%%%%%%
\begin{figure*}[ht!]
\centering
\includegraphics[width=\columnwidth]{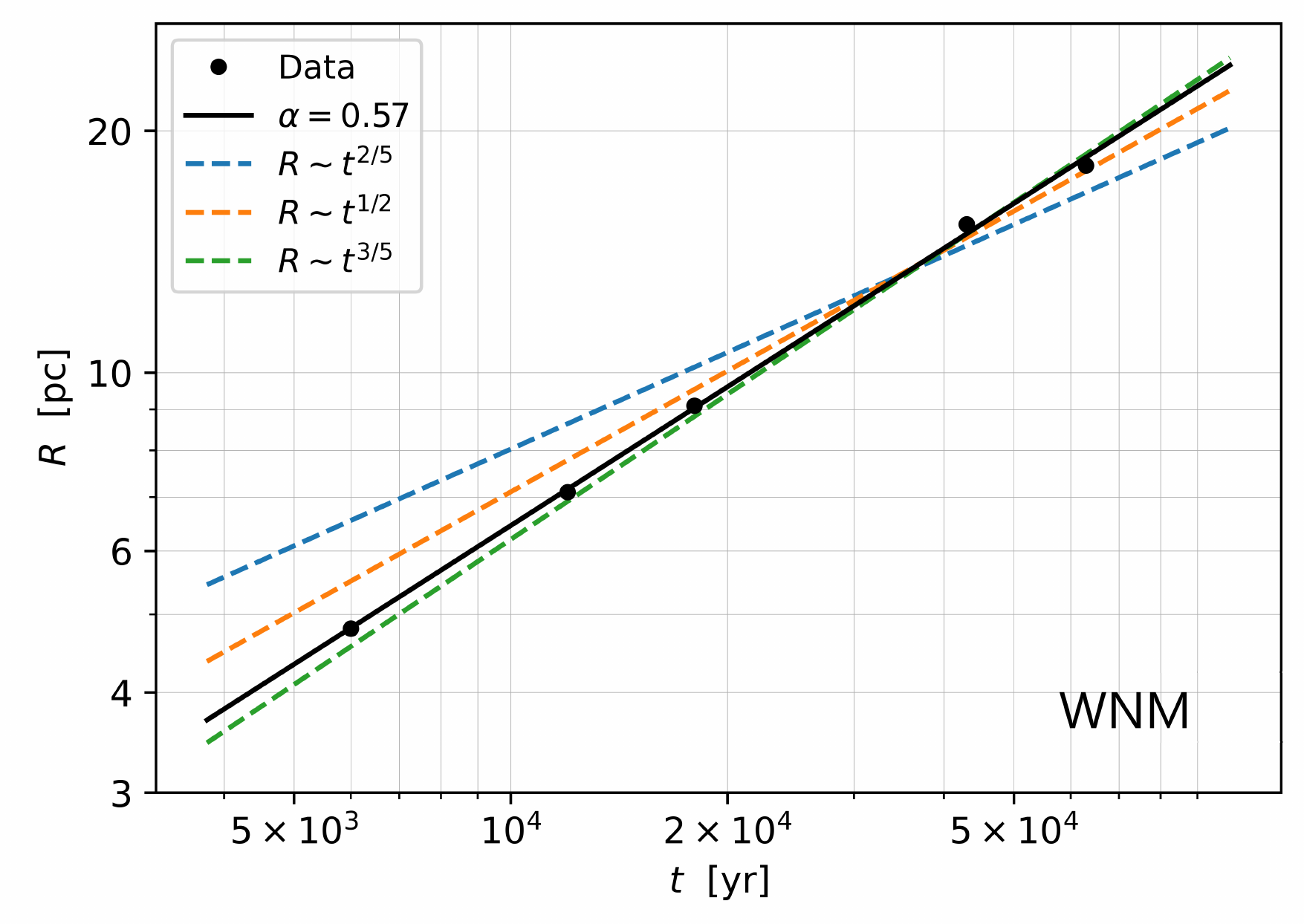}
\includegraphics[width=\columnwidth]{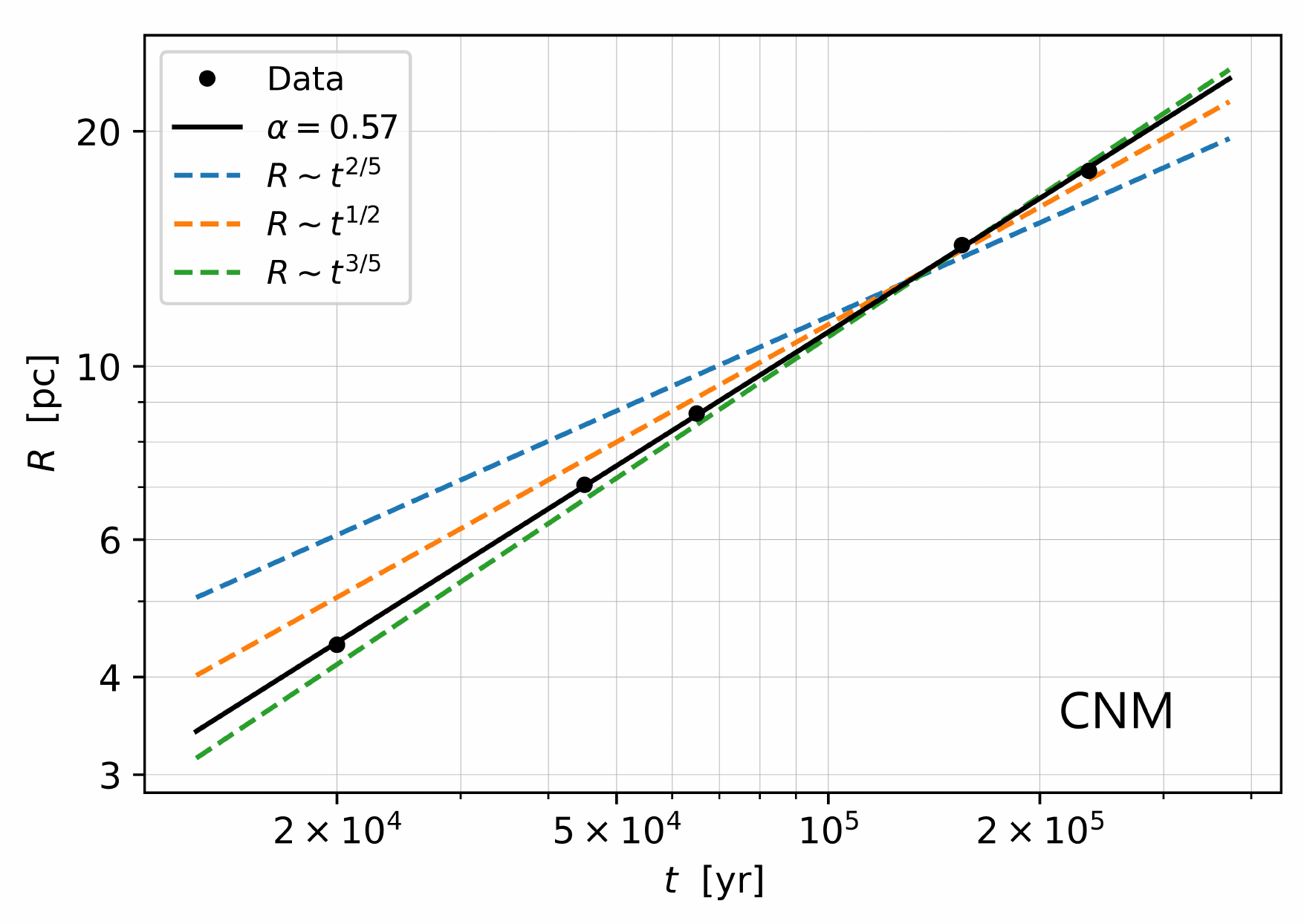}
\vspace{-2mm}
\caption{Comparison of the bubble expansion dynamics $R(t)$ with well-known theoretical expansion laws. \textit{Left:} WNM case. \textit{Right:} CNM case. Black points show the growth of the bubble's average forward shock radius over time $t_{\rm evol}$. Three specific times are shown as magnetic field slices for both ISM cases in Fig.~\ref{fig5}. A black straight line approximates the data with a power law $R \propto t^{\alpha}$. For comparison, colored dashed lines show well-known analytic expansion laws (see, e.g., Sedov 1946; Weaver et al. 1977).}
\label{fig3}
\end{figure*}
%%%%%%%%%%%%%%%%%%%%%%%%%%%%%%%%%%%%%%%%%%%%%%%%%%%%%%%%%
%%%%%%%%%%%%%%%%%%%%%%%%%%%%%%%%%%%%%%%%%%%%%%%%%%%%%%%%%
\begin{figure*}[ht!]
\centering
\includegraphics[scale=0.6]{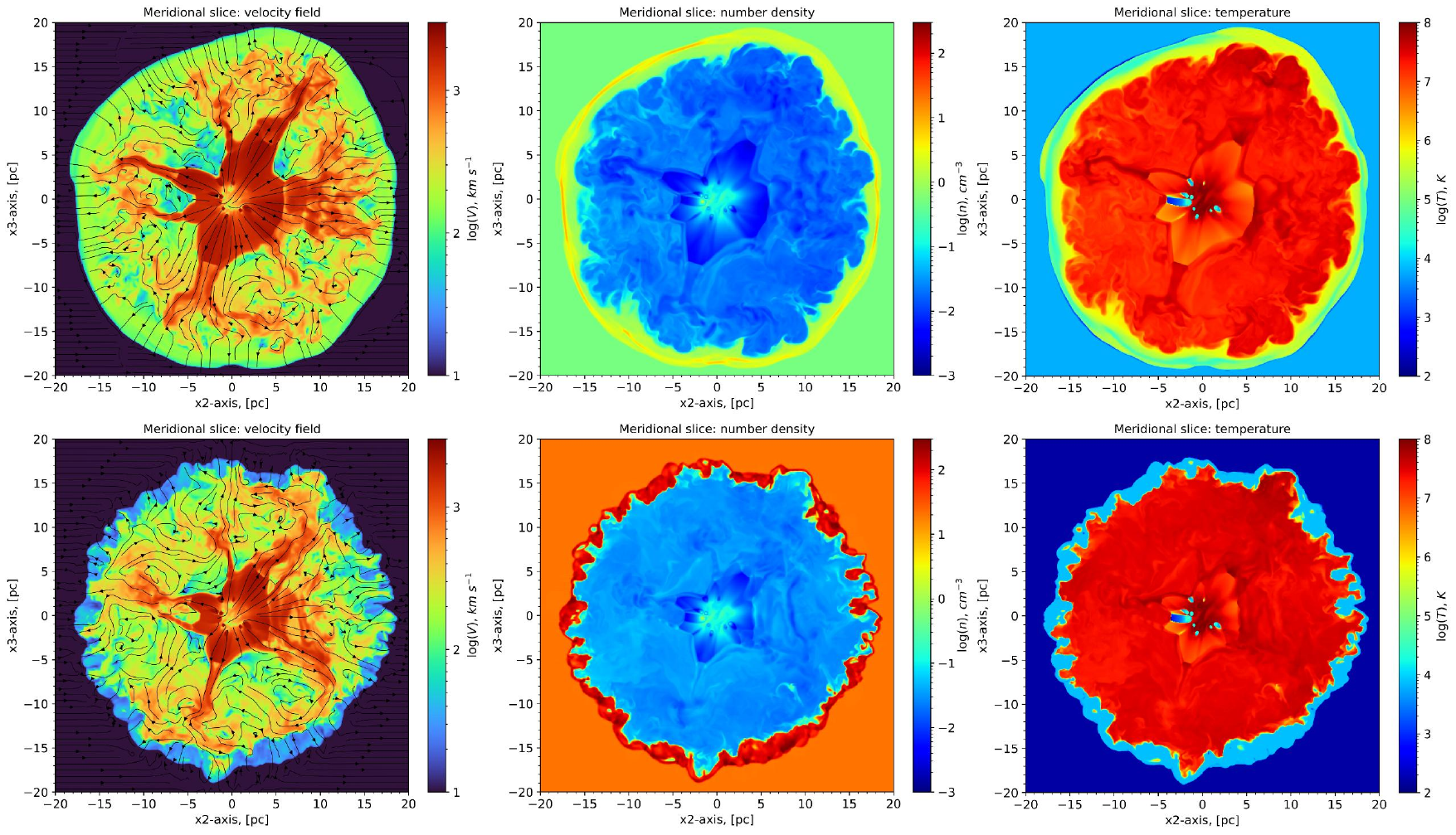}
\vspace{-1mm}
\caption{Comparison of physical quantities in the bubble, shown in a central $x_2$--$x_3$ plane slice. \textit{Top row:} the WNM case at $t_{\rm evol}=63$ kyr. \textit{Bottom row:} the CNM case at $t_{\rm evol}=235$ kyr. Each row displays, from left to right, maps of the velocity field, density, and temperature. Black arrows on the velocity maps indicate the local flow direction.}
\label{fig4}
\end{figure*}
%%%%%%%%%%%%%%%%%%%%%%%%%%%%%%%%%%%%%%%%%%%%%%%%%%%%%%%%%

%------------------------------------------------------
\section{Results}
\subsection{Cluster Wind Bubble Expansion Dynamics}
The interaction of the wind from a compact massive star cluster with the ISM leads to the formation of an expanding bubble. We estimated a power-law expansion, $R(t)=At^{\alpha}$, for this bubble in two ISM types (WNM and CNM), based on measurements of its outer shell radius at five evolutionary times $t_{\rm model}$ using the least-squares method. Fig.~\ref{fig3} presents the fitting results, yielding $A_{\rm wnm}\approx3.3\times10^{-2}$ pc yr$^{-\alpha}$, $A_{\rm cnm}\approx1.6\times10^{-2}$ pc yr$^{-\alpha}$, with $\alpha\approx0.57$ in both cases. The derived expansion law, $R \propto t^{0.57}$, lies between two well-known theoretical solutions: $R \propto t^{0.6}$ for an adiabatic (energy-driven) bubble and $R \propto t^{0.5}$ for a radiative (momentum-driven) bubble (see Avedisova, 1972; Arthur, 2007). We also note its proximity to the $R \propto t^{0.58}$ law obtained in the classic work by Weaver et al. (1977) when including radiative losses behind the wind termination shock front at a level $L_{\rm b}\lesssim\dot{E}_{\rm tot}$.

Figure 4 shows maps of the velocity field, density, and temperature for the WNM and CNM cases. When the shell reaches a radius of $\sim20$ pc, its expansion velocity is $\sim100-150$\kms (Mach $1-5$) for the WNM, while for the denser CNM case the velocity drops to $30-50$\kms (Mach $2-3$). The velocity field maps also reveal the cluster wind termination shock front, where the velocity drops sharply by an order of magnitude from $\sim2000$\kms to $\sim100-200$\kms. This front has an irregular surface morphology due to the random stellar distribution within the YMSC core, and it locally extends to distances $\lesssim5$ pc from the cluster center. At the extremities of the most distant parts of this shock, narrowing recollimation flows are visible, with velocities reaching up to $\sim1000$\kms. The density maps clearly show a structural difference between the two bubble shells. In the CNM case, the shell density reaches $\sim300$ cm$^{-3}$ (corresponding to $\sim10^4$\Msun\,swept up over $t_{\rm model}=235$ kyr) and is unstable (see~\S\ref{sec:dis}). The temperature within the bubble shell is about $8300$~K (the equilibrium temperature for ionized gas; see~\S\ref{sec:cool}) for the CNM, and $\sim10^4-10^5$~K for the WNM. In the latter case, the temperature spread is associated with the onset of the radiative phase in the shell. The bubble's interior, at densities $\gtrsim0.01$~cm$^{-3}$, is heated relatively uniformly to X-ray temperatures of $\gtrsim10^7$~K.

\subsection{Morphology of the Amplified Magnetic Field in the Shell}
%%%%%%%%%%%%%%%%%%%%%%%%%%%%%%%%%%%%%%%%%%%%%%%%%%%%%%%%%
\begin{figure*}[ht!]
\centering
\includegraphics[scale=0.6]{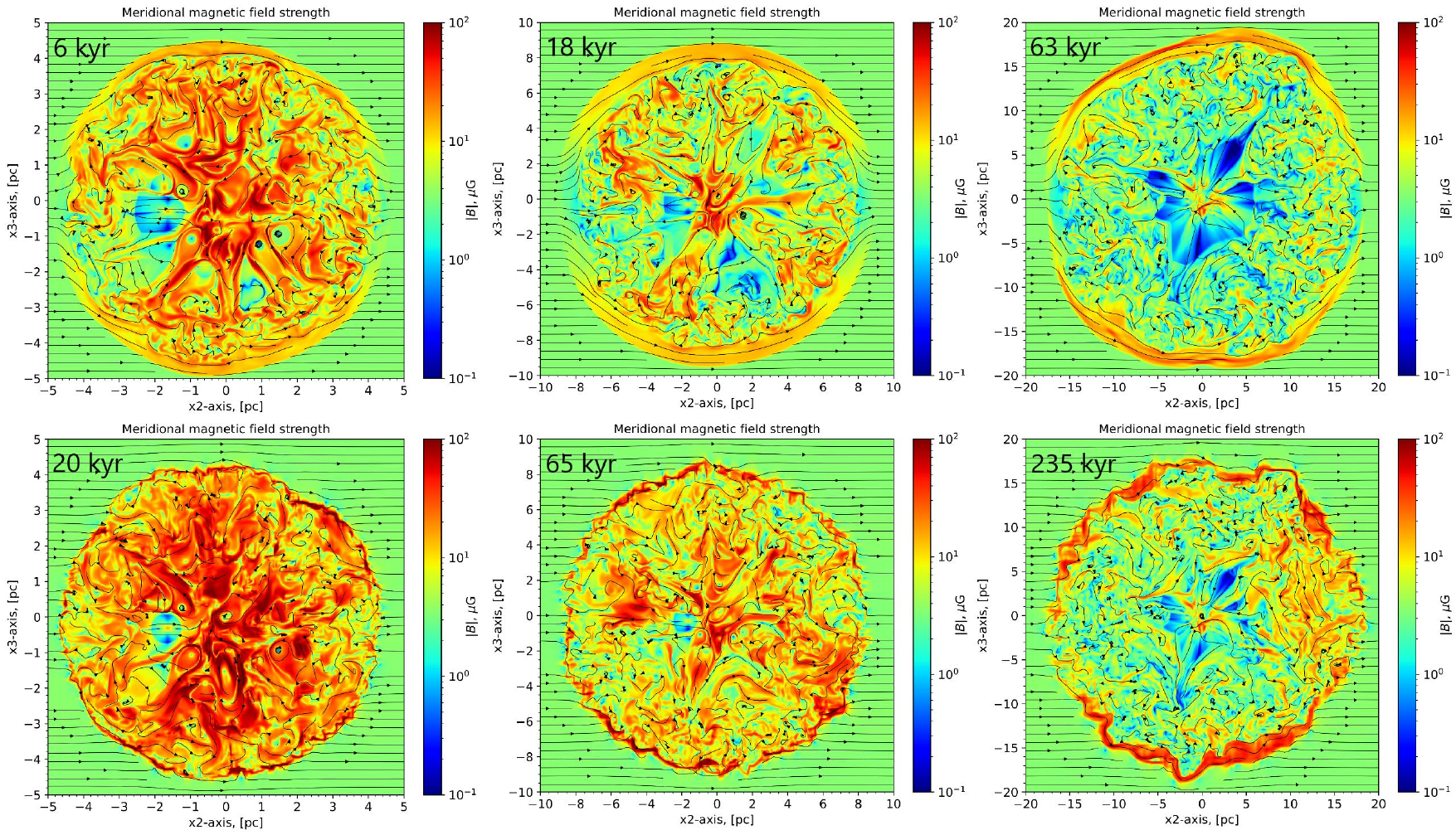}
\vspace{-1mm}
\caption{Comparison of magnetic field maps in a central $x_2$--$x_3$ plane slice of the bubble. \textit{Top row:} the WNM case. \textit{Bottom row:} the CNM case. Each row displays the field strength at three different evolutionary times $t_{\rm evol}$, ordered chronologically from left to right as the simulation domain is rescaled from 5 to 10 to 20 pc. Black arrows show the local magnetic field direction.}
\label{fig5}
\end{figure*}
%%%%%%%%%%%%%%%%%%%%%%%%%%%%%%%%%%%%%%%%%%%%%%%%%%%%%%%%%
\begin{figure*}[ht!]
\centering
\includegraphics[width=\columnwidth]{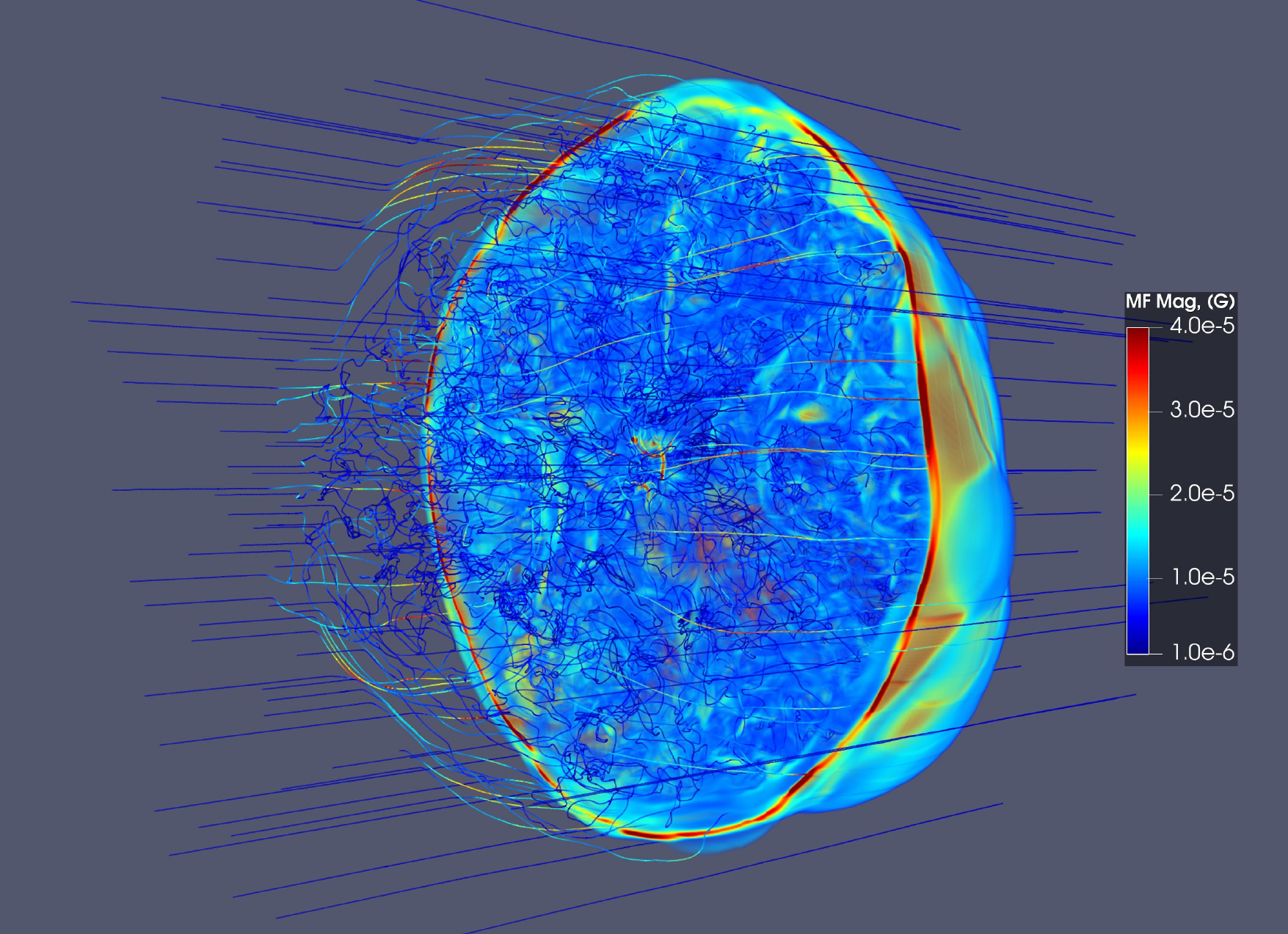}
\includegraphics[width=\columnwidth]{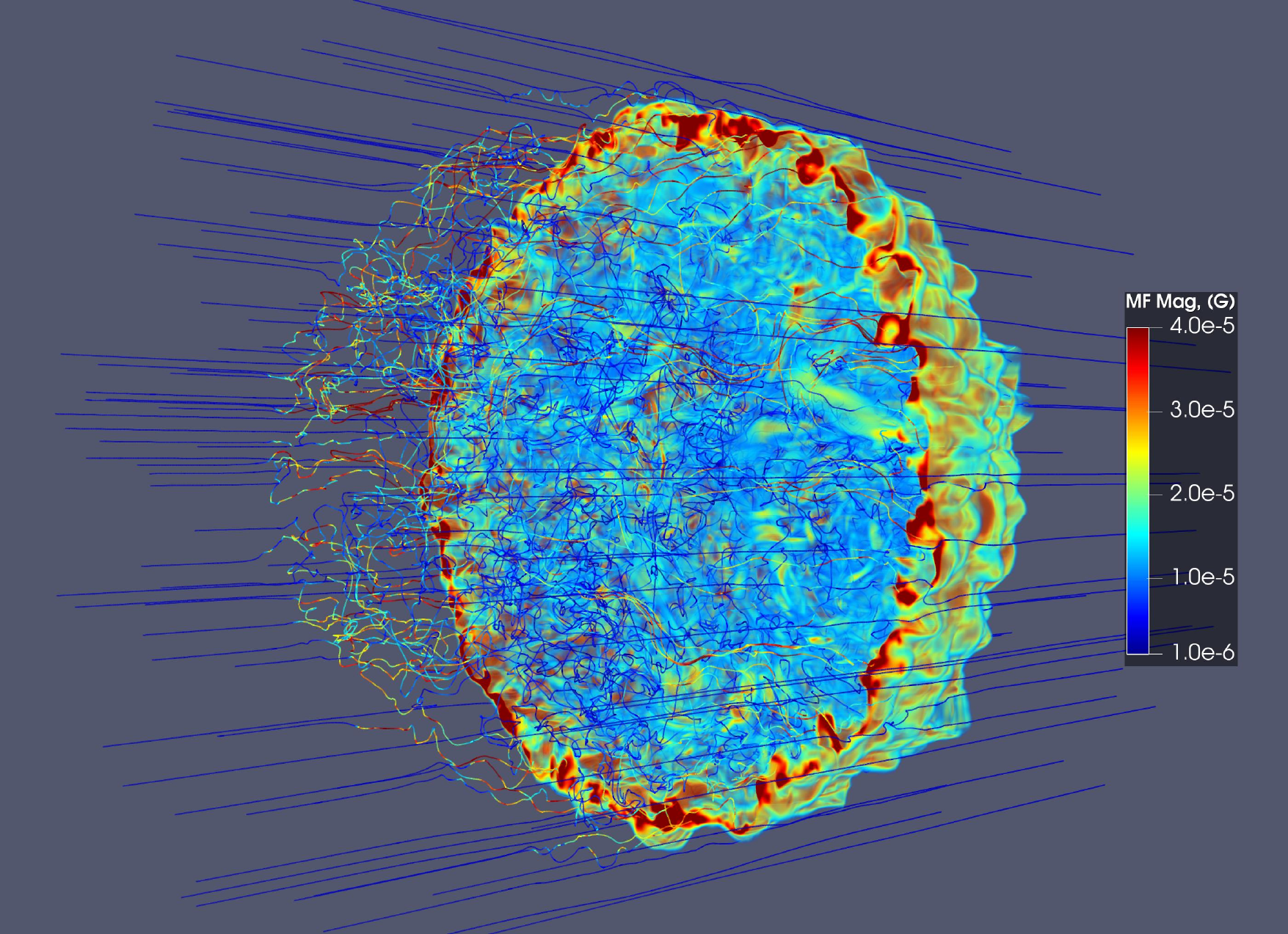}
\caption{Three-dimensional magnetic field structure of the cluster wind bubble in a central $x_1$--$x_2$ plane cross-section. \textit{Left:} the WNM case at $t_{\rm model}=63$ kyr. \textit{Right:} the CNM case at $t_{\rm model}=235$ kyr. Magnetic field lines are rendered as threads, outlining the bubble's structure in the cut-away region. For clarity, the color scale is artificially saturated at 40~$\mu$G; field strengths exceeding this value are confined to the densest gas clumps within the shell (see magnetic field slices in Figs.~\ref{fig4} and \ref{fig5}).}
\label{fig6}
\end{figure*}
%%%%%%%%%%%%%%%%%%%%%%%%%%%%%%%%%%%%%%%%%%%%%%%%%%%%%%%%%
\begin{figure*}[ht!]
\centering
\includegraphics[scale=0.6]{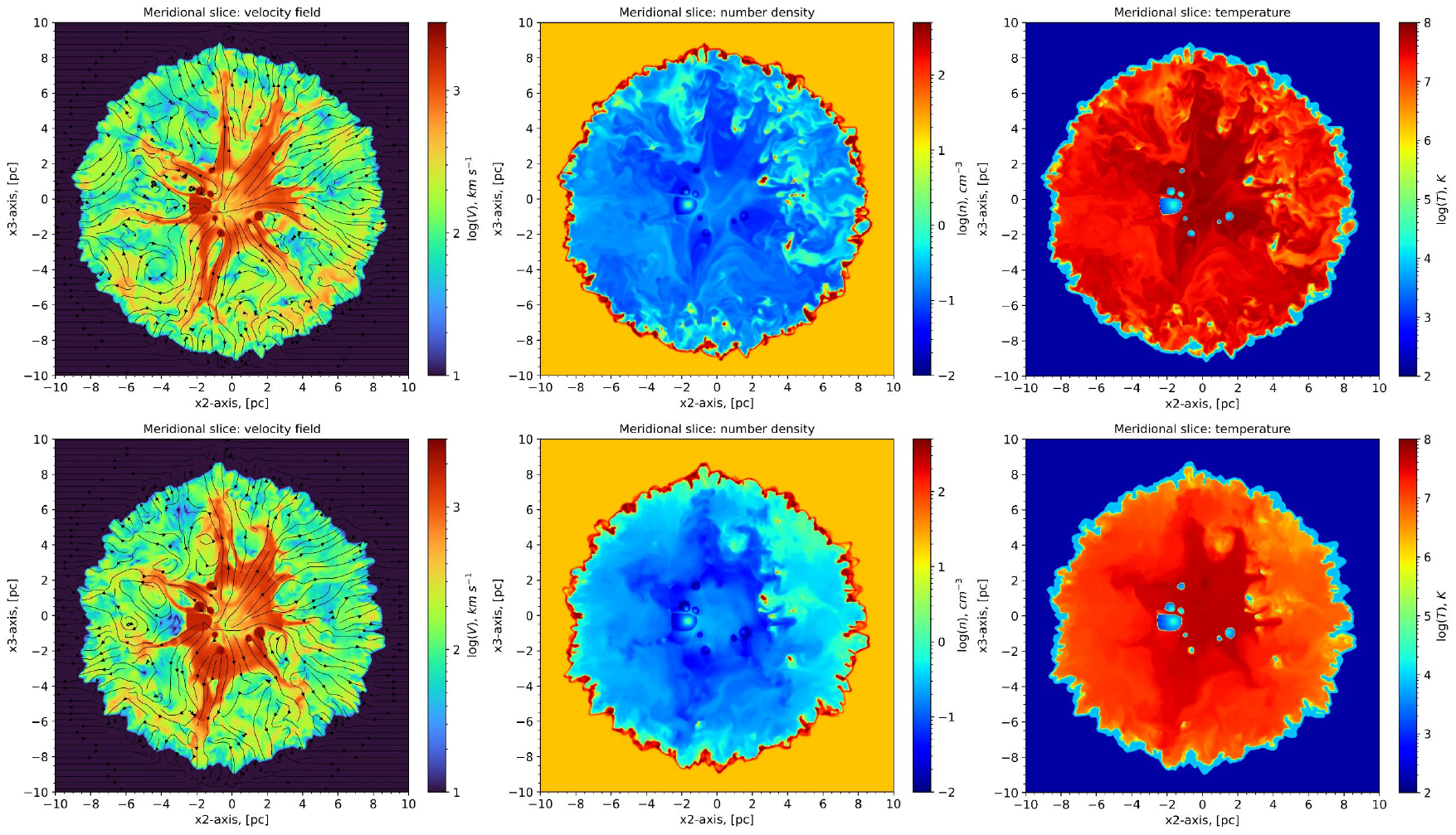}
\vspace{-1mm}
\caption{Comparison of simulation results for the cluster wind bubble expanding into the CNM at $t_{\rm model}=65$ kyr (i.e. 2nd scaling step: 10~pc). \textit{Top row:} without thermal conduction. \textit{Bottom row:} with thermal conduction. Each row shows, from left to right, maps of the velocity field, density, and temperature. Black arrows on the velocity maps indicate the local flow direction.}
\label{fig7}
\end{figure*}
%%%%%%%%%%%%%%%%%%%%%%%%%%%%%%%%%%%%%%%%%%%%%%%%%%%%%%%%%

Fig.~\ref{fig5} shows magnetic field maps at successive evolutionary times $t_{\rm model}$ for the two considered ISM phases, WNM and CNM. The magnetic field within the bubbles is highly turbulent throughout the entire simulation. At the initial stage, the field strength peaks at $\sim100$ $\mu$G in the YMSC core region, then gradually declines as the bubble expands and the wind termination shock develops. Subsequently, the bubble shell becomes the most magnetized region, where, on scales of $\sim20$ pc, the field is amplified in zones of gas compression (condensation) to $\sim10-50$ $\mu$G for the WNM and to $\sim50-70$ $\mu$G for the CNM. Inside the bubble, downstream of the termination shock front, the turbulent field varies in strength between $\sim1-20$ $\mu$G. The amplification in the shell is driven by thermal condensation of the gas (likely via an instability, see Section~\ref{sec:dis}), which is particularly effective at the higher ambient density of the CNM ($n_{\rm ism} \sim 10$ cm$^{-3}$). In the WNM case, the shell remains stable at least until $t_{\text{model}} = 63$ kyr. A comparison of the maps at $t_{\rm model}=18$ and $63$ kyr reveals the onset of active thermal condensation: the shell's forward shock becomes radiative, leading to shell thinning and magnetic field amplification in the compression regions. For the CNM case shown in the lower panels, the magnetic field in the shell is amplified to $\sim100$ $\mu$G when the bubble radius is 10 pc. It subsequently decreases to $\sim50-70$ $\mu$G as the overcooled shell gas is reheated to 8300 K by hydrogen photoionization from the YMSC radiation field. The uniformity of the ambient magnetic field leads to the formation of <<magnetic holes>> in the shell, where the external field lines are oriented perpendicular to its surface. Fig.~\ref{fig6} presents the three-dimensional magnetic field structure in a central cross-section of the bubble. In the CNM case, the instability generates a distinct cellular structure of the shell. Inside the dense cells, the magnetic field is amplified, reaching strengths up to 70 $\mu$G. The magnetic field lines clearly delineate the stable (WNM) and unstable (CNM) shell morphologies.

\subsection{Thermal Conduction Effects}
Including thermal conduction into the 3D MHD model of the cluster wind bubble may increase the computational cost by nearly two orders of magnitude. However, when this process is efficient ($t_{\rm tc} < t_{\rm model}$), it can produce several significant physical effects. Therefore, we included thermal conduction (see \S~\ref{sec:tc}) only for the first scaling step (from 5 to 10 pc) in the simulation of a bubble expanding into the CNM, which develops a dense shell. Fig.~\ref{fig7} compares the CNM bubble simulations with and without thermal conduction. With conduction included, the bubble is somewhat smaller, and its expansion follows a power-law closer to $R \propto t^{0.55}$ (cf. Fig.~\ref{fig3}). The velocity maps show that the forward shock expands at a slightly lower velocity of $\sim30-50$\kms, compared to the broader range of $50-100$\kms in the non-conductive case.

The most prominent differences are visible in the density and temperature maps. With thermal conduction, the density map appears significantly brighter on the color scale, indicating an approximately three times higher average density of $\sim0.15$ cm$^{-3}$ (with a range of $\sim0.1-0.3$ cm$^{-3}$) compared to the simulation without conduction. The density increases toward the dense, cold bubble shell, suggesting that the shell is undergoing evaporation. Thermal conduction makes the bubble's interior more uniform: gradients are smoothed, and small-scale fluctuations are suppressed. A similar qualitative result is seen in the temperature maps. The average temperature drops by about a factor of two, to $\sim10^7$ K, due to the inflow of cold, evaporating gas from the shell, which notably cools the region near the shell's inner edge.

\section{Discussion}\label{sec:dis}

\subsection{Instabilities in the Bubble Shell}
In the dense interstellar medium (CNM), the bubble shell was found to be unstable to fragmentation on timescales shorter than the simulation time $t_{\rm model}$. The dynamics of a radiative shell can involve the development of instabilities. The criteria for thermal instability were established by Field (1965) and Balbus (1986), while dynamical instabilities of a thin shell were investigated by Vishniac (1983) and Vishniac \& Ryu (1989). Models incorporating the potential development of both instabilities have also been discussed (see, e.g., Schure et al. 2009; Krause et al. 2013; Minier et al. 2018; Badjin et al. 2016, 2021).

\subsubsection{Thermal Instability}
The characteristic timescale of radiative processes in the shell is determined by the relation
\begin{equation}
    t_{\rm cool}=\frac{p}{(\gamma-1)n^2\Lambda(T)}.
\end{equation}
In general, cooling can effectively condense gas if the cooling time $t_{\rm cool}$ is shorter than the shell's dynamical crossing time, $t_{\rm dyn}=\delta R_{\rm sh}/v_{\rm sh}$. A detailed analysis of thermal instability in radiative shocks within supernova remnants was presented by Badjin et al. (2021). In their MHD simulations for temperatures above $10^4$~K, the authors employed the cooling function of Schure et al. (2009) and calculated the effective temperature index $\beta_{\rm eff}\equiv\rm{d}\ln\Lambda(T)/\rm{d}\ln T$ (see their Fig.~1). For the case of isobaric thermal condensation, this index must satisfy the criterion $\beta_{\rm eff}<2$ (Sutherland et al. 2003). This $\beta_{\rm eff}$ criterion is met for temperatures $T\gtrsim2\times10^4$~K, except within a narrow window of $4\times10^4<T<7\times10^4$~K for the cooling function in the PIE regime (see \S~\ref{sec:cool}). Thermal instability can lead to shell fragmentation and the formation of a multiphase structure on scales $\lambda_{\rm F}<\lambda<\lambda_{\rm th}=c_{\rm s}t_{\rm cool}$, potentially mixing hot gas from the bubble interior with shell material. Under the stricter criterion $\beta_{\rm eff}<1$, thermal oscillations of the shell's shock front can occur (Chevalier \& Imamura 1982).

\subsubsection{Dynamical Instability of a Thin Shell}
For the onset of the dynamical instability (PDTSO) described by Vishniac (1983), Vishniac \& Ryu (1989), the thermally pressurized bubble expansion must slow down sufficiently ($R \propto t^{\alpha}, \, \alpha<0.8$), and the gas swept up by the bubble, effectively compressed due to rapid radiative losses, must form a thin shell ($\delta R \ll R$). For bubbles with a constant central energy source (wind), Vishniac \& Ryu (1989) computed a critical density jump, $\delta n\approx25$, between the unperturbed external medium and the shell, leading to instability. In the CNM case ($n_{\rm ism}=20$~cm$^{-3}$), this criterion is satisfied, as the shell density can locally reach values $n\sim600-700$~cm$^{-3}$.

Linear estimates for the characteristic growth time and wavelength of the fastest-growing ($\rm max$) mode of the instability (Vishniac \& Ryu, 1989) are:
\begin{equation}    \tau_{\rm{max}}\sim\Gamma_{\rm{max}}^{-1}=2\left(\frac{c_{\rm{s}}^{2}\sigma_{0}}{p_{i}\dot{V}_{\rm{s}}}\right)^{1/2},
\end{equation}
\begin{equation}
    \lambda_{\rm{max}}=2\pi c_{\rm{s}}^{2}\left(\frac{\sigma_{0}}{p_{i}\dot{V}_{\rm{s}}}\right)^{1/2},
\end{equation}
where $c_{\rm{s}}$ is the sound speed in the bubble shell, $p_{i}$ is the gas pressure within the bubble, $\dot{V}_{\rm{s}}$ is the shell deceleration, $\sigma_{0}$ is the column density through the shell (thickness). Using data from the CNM model: $c_{\rm{s}}^{2}\sim10^{12}$ cm$^{2}$~c$^{-2}$, $p_{i}\sim10^{-10}$ dyn~cm$^{-2}$, $\dot{V}_{\rm{s}}\sim10^{-6}$ cm~c$^{-2}$, $\sigma_{0}\sim10^{-4}$ g~cm$^{-2}$. It follows that $\lambda_{\rm{max}}\sim2$ pc and $\tau_{\rm{max}}\sim6\times10^{4}$ years.

\subsection{Compact Clusters and the Interstellar Medium}
Observational analysis of molecular cloud evolution (Schinnerer \& Leroy, 2024) suggests that massive star winds likely contribute to parent cloud destruction, occurring before the phase of multiple supernova explosions. We performed 3D MHD simulations of the bubble's expansion dynamics driven by the collective wind from a compact YMSC during its Wolf-Rayet-dominated phase, corresponding to an age $\lesssim 3$ Myr. In this work, the cluster's MHD plasma dynamics is modeled starting from the interaction of multiple individual massive star winds. Accounting for these individual wind interactions during the formation of the cluster wind is essential for studying young cluster feedback on its parent cloud (Polak et al. 2024).

Our single-fluid MHD model computes the plasma temperature $T$. Under conditions of collisionless temperature relaxation, the electron and ion temperatures would each be equal to $T/2$. In the central cluster core region, the plasma is heated to single-fluid temperatures $T\sim10$ keV. This initially high temperature of the thermalized cluster wind decreases during its supersonic expansion (see Fig.~\ref{fig4}). Over the lifetime of the Wolf-Rayet stars, $\sim10^5$ yr, the cluster wind thus forms a bubble roughly 20 pc in size, filled with hot X-ray plasma at temperatures $\sim1$ keV.

The results obtained can be of great importance for interpreting the X-ray observations of compact clusters. An outstanding Galactic example of a rich, compact YMSC likely approaching the age of its first supernovae is the Arches cluster. An analysis of its stellar population by Clark et al. (2018) revealed more than 10 WNLh stars among total 88 identified early type stars, yielding a cluster age estimate of $2-3.3$ Myr. This age implies that only stars with masses $\gtrsim120$\Msun could have exploded as supernovae. X-ray observations of the Arches cluster and its surroundings reveal a multi-component spectral structure with both thermal and non-thermal components (see, e.g., Krivonos et al. 2014; Clavel et al. 2014).

The location of the Arches and another notable compact YMSC, Quintuplet, in close proximity to the Galactic center with a high density of powerful transient and quasi-stationary sources in the Central Molecular Zone (CMZ) complicates the analysis of extended source observations (see a detailed analysis of the Galactic center region observations by Khabibullin et al. 2025). Along with X-ray emission from hot plasma in the cluster core and surrounding bubble, non-thermal emission from excitation of K-shell X-ray lines by low-energy cosmic rays is expected (see, e.g., Yusef-Zadeh et al. 2007), as well as synchrotron X-ray emission from electrons and positrons accelerated in the cluster to energies up to 100 TeV (see review, Bykov 2014). Furthermore, the propagation of X-ray emission fronts from powerful flares of the central black hole Sgr~A$\ast$ (Sunyaev et al. 1993) through CMZ clouds is accompanied by reflection from dense clouds, observed, in particular, in fluorescent iron K-shell lines as a non-stationary X-ray emission component.
%ALTERNATIVE: In addition to thermal X-ray emission from hot plasma in the cluster core and bubble, several non-thermal processes are expected. These include X-ray line excitation from low-energy cosmic rays (see, e.g., Yusef-Zadeh et al. 2007) and synchrotron X-ray emission from electrons and positrons accelerated in the cluster to energies up to 100 TeV (see review, Bykov 2014). Furthermore, X-ray emission fronts from powerful flares of the central black hole Sgr~A$\ast$ (Sunyaev et al. 1993) propagate through CMZ clouds, producing a variable reflection component observed in fluorescent iron K-shell lines.

Our modeling of the hot plasma and amplified magnetic fields in the interaction region between the ISM and the wind of a compact YMSC provides a basis for estimating the individual contributions of these components to the observed X-ray emission. For compact clusters within the Galactic CMZ, where the ambient ISM density and magnetic field strength are significantly higher than in our present model, a separate analysis will be required.

%------------------------------------------------------
\section{Conclusion}
\label{sect:concl}
This paper presents the results from 3D MHD simulations of plasma flow within a wind-driven bubble, formed in the interstellar medium by a compact cluster of massive stars prior to the first supernova explosions. At this evolutionary stage, a significant population of Wolf-Rayet stars with powerful stellar winds is present. The collision and partial thermalization of these winds generate a high-pressure region in the hot, compact cluster core. The expansion of this hot core plasma, together with the rest non-thermalized stellar wind flows, drives a collective cluster wind. Its interaction with the interstellar medium forms a dense expanding shell with an amplified magnetic field. In a cold neutral medium (CNM), instabilities can produce a distinctive cellular structure in the shell. During this Wolf-Rayet-dominated phase, the cluster wind can significantly reshape the surrounding parent cloud, sweeping over $10^4$ $M_{\odot}$ of gas into a fragmented shell before the first supernovae occur.

%bubble dynamics
Using the PLUTO code, we simulated the bubble's expansion into both a warm neutral medium (WNM) and a CNM until the shell reached $\sim20$ pc. The total cluster wind power was about $1.6\times10^{38}$\ergs. A key feature of this model is the direct MHD calculation of the partial stellar wind thermalization in the cluster core, which critically affects bubble formation. As a result, the expansion follows a power law $R(t)=At^{\alpha}$ with $\alpha\simeq0.57$, in agreement with theoretical predictions (Weaver et al. 1977). For both ISM types, an irregularly shaped termination shock forms, extending to radii $\gtrsim5$ pc. The bubble interior is heated to X-ray temperatures $\gtrsim10^7$ K.

%magnetic field
The magnetic field within the bubble is highly turbulent, reaching strengths $\gtrsim100$~$\mu$G near the young massive star cluster (YMSC) core at the earliest expansion stage. As the system evolves, the field amplifies within the compressed dense shell to $\sim10-50$~$\mu$G for the WNM and $\sim50-70$~$\mu$G for the CNM. So-called <<magnetic holes>> form in the regions of the shell where the ambient uniform magnetic field is perpendicular to its surface.

%thermal conduction
Including thermal conduction does not qualitatively alter the complex morphology of the unstable CNM bubble shell but does affect its dynamics, slightly decelerating the expansion. Notable temperature and density changes occur near the inner edge of the dense CNM shell due to its evaporation, while their sharp gradients in the bubble's interior are smoothed.

%instabilities
The shell morphology differs markedly between the two media. Expansion into the CNM produces a thin fragmented shell with a distinctive cellular morphology, while in the WNM the shell remains smooth and unfragmented. The CNM shell's cellular structure is likely caused by the combined action of thermal (Field 1965) and dynamical thin-shell instabilities (Vishniac 1983; Vishniac \& Ryu 1989).

%significance
These results are also relevant for interpreting observational data and disentangling the thermal and non-thermal components of X-ray emission from YMSCs on the verge of their first supernovae.

\section*{Acknowledgments}
The authors thank S.I. Blinnikov, D.A. Badjin, S.I. Glazyrin, and E.M. Urvachev for discussing the results and valuable comments. The authors thank A. Mignone and the PLUTO code development team. The work was supported by RSF grant \textnumero 25-72-20007. The modeling of thermal conduction effects was performed on the <<Tornado>> subsystem of the supercomputer center at Peter the Great St. Petersburg Polytechnic University.

\end{document}